\documentclass[aps,prd,onecolumn,superscriptaddress,nofootinbib,amsmath,amssymb]{revtex4}

\usepackage{graphicx}
\usepackage{enumerate}

 \tighten  \usepackage[hypertexnames=false]{hyperref}
\renewcommand{\d}{\partial}
\newcommand{\half}{\frac{1}{2}}

\renewcommand{\tilde}{\widetilde}

\newcommand{\bref}[1]{\textbf{\ref{#1}}}

\def\cE{\mathcal{E}}

\def\cJ{\mathcal{J}}

\def\cM{\mathcal{M}}

\def\cP{\mathcal{P}}

\def\cV{\mathcal{V}}
\def\cW{\mathcal{W}}

\begin{document}

 \title{Boundary dynamics of asymptotically flat 3D gravity coupled to higher spin fields.}

\author{Hern\'an A. Gonz\'alez}
\email{hgonzale-at-ulb.ac.be}
\thanks{Laurent Houart postdoctoral fellow.}
\affiliation{Physique Th\'eorique et Math\'ematique \\
Universit\'e Libre de Bruxelles \& International Solvay Institutes\\
Campus Plaine C.P. 231, B-1050 Bruxelles, Belgium.}

\author{Miguel Pino}
\email{miguel.pino.r-at-usach.cl}
\affiliation{Departamento de F\'isica, Universidad de Santiago de Chile, Av. Ecuador 3493, Estaci\'on Central, Santiago, Chile.}

\begin{abstract}
We construct a two-dimensional action principle invariant under a spin-three extension of BMS$_3$ group. 
Such a theory is obtained through a reduction of Chern-Simons action with a boundary. This procedure is carried out by
imposing a set of boundary conditions obtained from asymptotically flat spacetimes in three dimensions. When implementing part of this set, we obtain an analog of chiral WZW model based on a contraction of $sl(3,\mathbb{R}) \times sl(3,\mathbb{R})$.
 The remaining part of the boundary conditions imposes  constraints on the conserved currents of the model, 
 which allows to further reduce the action principle. 
 It is shown that a sector of this latter theory is related to a flat limit of Toda theory. 
\end{abstract}

\maketitle

\section{Introduction}

Three-dimensional gravity has proven to be an excellent laboratory for testing holographic dualities. 
For example, the study of asymptotically Anti-de Sitter (AdS) spacetimes was an early precursor of the AdS/CFT conjecture when Brown 
and Henneaux found that the group of symmetries at infinity correspond to the two-dimensional conformal group
 \cite{Brown:1986nw}. Furthermore, it was proved that Einstein-Hilbert action 
in presence of negative cosmological constant can be reduced to Liouville theory \cite{Coussaert:1995zp}, 
which is known to be conformally invariant. In this sense, Liouville theory can be regarded as a classical dual of 3D Einstein gravity with negative cosmological constant.

Recently, the interaction of higher spin fields with gravity 
has received a lot of attention. In particular, it has been noticed that AdS$_3$ asymptotics encodes not only
conformal symmetry, but also the infinite dimensional $W$-algebra, which involves excitations of higher spin 
fields \cite{Campoleoni:2010zq,Henneaux:2010xg}. In addition, as argued in \cite{Campoleoni:2010zq},
 three-dimensional higher spin gravity can be reduced to Toda theory, a generalization of Liouville model
 possessing higher spin conserved charges \cite{Bilal:1988jf,Bilal:1988ze}. This is a well-known fact due to 
 prior work of \cite{Forgacs:1989ac,Balog:1990mu} (for a review see e.g., \cite{Feher:1992yx}), where it was 
 established that Toda (Liouville) theory is obtained as a Hamiltonian reduction of Wess-Zumino-Witten (WZW) model.

When the cosmological term is absent, it is possible to find similar features as in the AdS$_3$ case.
In fact, three-dimensional asymptotically flat spacetimes contain an infinite set of transformations
which preserves the conformal structure of null infinity \cite{Barnich:2006av,Ashtekar:1996cd}, which gives rise to
the so-called $bms_3$ algebra, whose representation through Poisson brackets picks up a central term
\cite{Barnich:2006av}. These set of transformations are in correspondence to the AdS case when one takes a suitable vanishing cosmological constant limit \cite{Barnich:2012aw}.
 Related recent work using this approach, following the holographic renormalization program, appeared in \cite{Costa:2013vza,Fareghbal:2013ifa}.  In \cite{Krishnan:2013wta}, 
a map between flat and AdS$_3$ geometries was developed by introducing Grassmann variables. In the context of non-relativistic two-dimensional systems, it has been shown that $bms_3$ algebra is isomorphic to Galilean Conformal Algebra, $gca_2$ \cite{Bagchi:2010zz,Bagchi:2012cy}. 
 
Interestingly enough, higher spin gravity in three dimensions on a flat background can be formulated in terms of a
Chern-Simons action \cite{Blencowe:1988gj}. In this setup, higher spin fields can be consistently truncated to a finite set.
Moreover, a set of asymptotic conditions can be devised in order to
accommodate a higher spin generalization of  $bms_3$ algebra \cite{Gonzalez:2013oaa,Afshar:2013vka}. 
These results were re-obtained from the ones found for asymptotically AdS spacetimes by using a gauge choice 
that allows to take the flat spacetime limit.

In this note we use the asymptotic conditions presented in \cite{Gonzalez:2013oaa,Afshar:2013vka} in 
order to reduce Chern-Simons action down to a first order Lagrangian involving scalar fields. This model is a generalization of the $bms_3$-invariant theory found in \cite{Barnich:2012rz} and it can be related to a flat spacetime limit of $sl(3,\mathbb{R})$ Toda theory. Furthermore, this reduction will also be studied at the level of the conserved quantities present in the theory and their algebra. 

The paper is organized as follows. In the next section, specific sets of the asymptotic conditions are selected from the on-shell gauge field. It is also explained how they will be
used to reduce the theory.  

Section \bref{CSWZW} is devoted to the construction of a well defined Chern-Simons action principle. With this at hand,  we solve the constraints inside of this regularized action, producing a two-dimensional theory, analog to a WZW model \cite{Witten:1983ar} based on a contraction of 
$sl(3,\mathbb{R}) \times sl(3,\mathbb{R})$.

In section \bref{symWZW}, the symmetries of this latter model are reviewed. It is shown to possess two types of conserved currents, 
whose Poisson brackets give rise to a Kac-Moody algebra.

In section \bref{sec:2ndbms}, we build charge densities as bilinear and cubic invariants of the currents. 
These cubic quantities are suitable generalizations of the Sugawara construction\footnote{In the conformal case, 
cubic invariants were developed in \cite{Bais:1987dc}.}. The algebra of these densities produces a 
spin-three extension of $bms_3$ algebra without central extensions.

Part of the boundary conditions discussed in section \bref{CB} will impose constraints on the conserved currents of the WZW model. In section \bref{RA} these conditions will be first used to further reduce the theory at the level of the action. There, we make use of the Gauss decomposition for the group element and conveniently rewrite the constraints in order to impose them directly inside of the Lagrangian. The outcome of this procedure is a theory which can be regarded as the classical dual of asymptotically flat gravity coupled to spin-three fields\footnote{For recent works on flat space holography, see \cite{Bagchi:2012yk,Bagchi:2012xr,Barnich:2012xq,Bagchi:2013qva,Bagchi:2013lma,Detournay:2014fva}.}. Moreover, as shown in section \bref{Toda}, a sector of this reduced theory is related by a field redefinition with a flat limit of Toda theory written in Hamiltonian form.  In the second part of section \bref{RA}, we implement further conditions on the currents  which enables to define Dirac brackets of the reduced theory. Under these brackets, linear and central terms are added to the Poisson algebra computed in the previous section.  As expected, this result coincides exactly with the previous analysis done in \cite{Gonzalez:2013oaa,Afshar:2013vka}. 

Finally, in section \bref{higher}, we comment on the generalization of these results to higher rank algebras.   


\section{Boundary conditions}
\label{CB}

Our starting point is the Chern-Simons formulation in (2+1)-dimensions. For simplicitly, we will work with the spin-three case, closely following the conventions of \cite{Gonzalez:2013oaa}.
In particular, we consider a gauge field $A$, 
\begin{equation}
A=\omega ^{a}J_{a}+e^{a}P_{a}+W^{ab}J_{ab}+E^{ab}P_{ab}\ ,
\end{equation}
where $J_a$ and $P_a$ comprise the Poincar\'e algebra whereas $J_{ab}$ and $P_{ab}$ are new symmetric and traceless generators\footnote{In the sense that $\eta^{ab}J_{ab}=\eta^{ab}P_{ab}=0$.}. All of them satisfy the following commutation relations,
\begin{equation}
\label{algs3}
\begin{split}
& [J_{a},J_{b}]=\epsilon _{abc}J^{c},\quad {}[P_{a},J_{b}]=\epsilon
_{abc}P^{c},\quad {}[P_{a},P_{b}]=0, \\
&{}[J_{a},J_{bc}]=\epsilon _{\;\;a(b}^{m}J_{c)m},\quad
[J_{a},P_{bc}]=\epsilon _{\;\;a(b}^{m}P_{c)m}, \quad {}[P_{a},P_{bc}]=0,\\
& {}[J_{ab},J_{cd}]=-\left( \eta _{a(c}\epsilon _{d)bm}+\eta _{b(c}\epsilon
_{d)am}\right) J^{m},\quad {}[J_{ab},P_{cd}]=-\left( \eta _{a(c}\epsilon
_{d)bm}+\eta _{b(c}\epsilon _{d)am}\right) P^{m}, \quad {}[P_{ab},P_{cd}]=0,
\end{split}
\end{equation}%
This algebra has been obtained as a Wigner-In\"{o}n\"{u} contraction of $sl(3,\mathbb{R})\times sl(3,\mathbb{R})$. See appendix \bref{a1} for conventions.

As it was shown in \cite{Gonzalez:2013oaa,Afshar:2013vka}, it is possible to find a class of solutions to the equations of motions which possess the following form:
\begin{multline}
\label{Aonshell}
A=\left( \frac{1}{2}\mathcal{M}du-dr+\left( \mathcal{J}+\frac{u}{2}\d_{\phi }\mathcal{M}\right) d\phi \right) P_{0}+duP_{1}+rd\phi P_{2}+\frac{1}{
2}\mathcal{M}d\phi J_{0}+d\phi J_{1} \\+\left( \mathcal{W}du+\left( \mathcal{V}+u\partial _{\phi }\mathcal{W}
\right) d\phi \right) P_{00}+\mathcal{W}d\phi J_{00},
\end{multline}
where $r$ is a radial coordinate, $\phi$ is an angle and $u$ is a null coordinate playing the role of time. Here $\cM$, $\cJ$, $\cW$ and $\cV$ stand for arbitrary functions of $\phi$. Under $bms_3$ transformations $\mathcal{M}$, $\mathcal{J}$ transform as spin-two generators, while $\mathcal{W}$, $\mathcal{V}$ do as spin-three.  

The above solution encodes all the necessary information to reduce CS action to a theory defined on the boundary of the spacetime. We will take into account some features of \eqref{Aonshell} in an off-shell formulation. In particular, it will be important to consider:
\begin{enumerate} [(i)]
\item In order to define a differentiable action principle (in the sense of the Regge-Teitelboim approach, \cite{Regge:1974zd}) we will implement,
\begin{equation*}
\omega_{u}^{a}=0,\: e_{u}^{a}=\omega_{\phi}^{a},\quad W_{u}^{ab}=0,\: E_{u}^{ab}=W_{\phi}^{ab},
\end{equation*} \label{i}
as boundary conditions. This allows to reduce Chern-Simon action down to a flat version of WZW model. 
\item The following set of conditions will impose first class constraints on certain components of the conserved currents associated to the flat WZW model. In turn, they further reduce the model obtained using condition (i). They are given by
\begin{equation*}
 e^1_\phi=0,\: \omega^1_\phi=1,\quad E^{11}_\phi=E^{12}_{\phi}=W^{11}_\phi=W^{12}_{\phi}=0.
\end{equation*} \label{ii}
\item  When constructing Dirac brackets, we will impose extra constraints in order to completely fix the gauge freedom. These constraints can be read off from
\begin{equation*}
e^2_\phi=r,\: \omega^2_\phi=0, \quad E^{01}_\phi=E^{02}_{\phi}=W^{01}_\phi=W^{02}_{\phi}=0.
\end{equation*} \label{iii}
\end{enumerate}
In the forthcoming sections we will use each of these conditions to perform every step of the reduction.
The analysis considers a boundary at future null infinity, $r \to \infty$ and $u=\text{const}$. Also, we do not take into account either contributions of localized sources or global aspect such as holonomies.

\section{From CS to WZW}
\label{CSWZW}

To begin with, let us consider Chern-Simons theory in three dimensions, in presence of a boundary $\d \cM$,
\begin{equation}
I[A]=\frac{k}{4\pi }\int_{\cM} \langle AdA+\frac{2}{3}A^{3}\rangle \ ,  \label{CS2}
\end{equation}%
where the manifold $\cM$ is considered to be the real line times the disc. The bracket $\langle \cdot \cdot \cdot \rangle $ in \eqref{CS2} stands for a non-degenerate invariant bilinear product (see \cite{Gonzalez:2013oaa} for details). The Newton constant is related to $k$ by
\begin{equation}
k=\frac{1}{4G}.
\end{equation}

Due to its topological nature, it has been realized that this theory can be defined as an action over $\d \cM$ \cite{Witten:1988hf,Elitzur:1989nr}. 
In order to show this fact for Chern-Simons based on Lie-algebra \eqref{algs3}, it will be convenient to express action \eqref{CS2} in terms of generalized driebein and spin connection 
\begin{align}
\cE &= e^aJ_a+E^{ab}J_{ab},\\
\Omega&=\omega^aJ_a+W^{ab}J_{ab},
\end{align}
where $J_a$ and $J_{ab}$ form a representation of $sl(3,\mathbb{R})$
\begin{subequations}
\label{sl3}
\begin{eqnarray}
[J_{a},J_{b}]&=&\epsilon _{abc}J^{c},\\
{}[J_{a},J_{bc}]&=&\epsilon _{\;\;a(b}^{m}J_{c)m},\\
{}[J_{ab},J_{cd}]&=&-\left( \eta _{a(c}\epsilon _{d)bm}+\eta _{b(c}\epsilon_{d)am}\right) J^{m}.
\end{eqnarray}
\end{subequations}
These variables transform action \eqref{CS2} into an equivalent Einstein-Hilbert form 
\begin{equation}
I=\frac{k}{\pi}\int {\rm Tr}\left[\cE (d\Omega+\Omega^2)\right], \label{EHform}
\end{equation}
where the non-vanishing traces are given by
\begin{subequations}
\label{traza}
\begin{align}
{\rm Tr}[J_a J_b]&=\frac{1}{2}\eta_{ab},\\ 
{\rm Tr}[J_{ab} J_{cd}]&=\frac{1}{2}(\eta_{ac}\eta_{bd}+\eta_{ad}\eta_{cb}-\frac{2}{3}\eta_{ab}\eta_{cd}).
\end{align}
\end{subequations}
Reducing \eqref{EHform} to a boundary theory must be done in two steps. First, a well-defined action principle is in order, in which the classical set of solutions \eqref{Aonshell} constitute a true extremum, in the sense that
\begin{align}\label{extremo}
\delta I_{\rm on-shell}=0.
\end{align}
This is accomplished by adding boundary terms to the action such that the asymptotic behavior of the fields \eqref{Aonshell} fulfills \eqref{extremo}. Secondly, the constraints must be solved. Both steps can be easily done in the Hamiltonian formulation. By performing a decomposition in time and space, i.e. $\cE=\cE_u du + \cE_i dx^i$ and $\Omega=\Omega_u du + \Omega_i dx^i$, one can see that the action is already written in Hamiltonian form after discarding boundary terms
\begin{align}
I=\frac{k}{\pi}\int d^3x \: \epsilon^{ij} {\rm Tr}\left[\cE_i\dot{\Omega}_j+\cE_u(\d_i\Omega_j+\Omega_i\Omega_j)+\Omega_u(\d_i\cE_j+\Omega_i\cE_j+\cE_i\Omega_j)\right].
\label{Haction}
\end{align}
If we the take the variation of $I$, we obtain
\begin{align}
\delta I= {\rm ``Bulk}\:{\rm piece"}+\frac{k}{\pi}\int_{r = \infty}  du d\phi\; {\rm Tr}\left[\cE_u\delta \Omega_\phi + \Omega_u \delta \Omega_\phi\right],
\end{align}
Note that when evaluated on-shell,  the bulk piece vanishes. The remaining boundary term becomes integrable when the boundary conditions \eqref{i} are taken into account. In terms of  $\cE$ and $\Omega$ variables, these conditions read 
\begin{align}
\Omega_{u}=0, \quad \cE_{u}=\Omega_{\phi},
\end{align}
with which one can define a new action principle given by,
\begin{equation}
\label{WdAc}
\tilde{I}=I-\frac{k}{2\pi}\int  du d\phi\; {\rm Tr}\left[\Omega_\phi^2\right],
\end{equation}
that satisfy \eqref{extremo}. Now we are able to solve the constraint and use them inside of $\tilde{I}$. The constraints are
\begin{align}
\epsilon^{ij}(\d_i\Omega_j+\Omega_i\Omega_j) = 0,\\
\epsilon^{ij}(\d_i\cE_j+\Omega_i\cE_j+\cE_i\Omega_j) = 0,
\end{align}
and can be locally solved, yielding\footnote{To obtain \eqref{eqe}, note that using $\Omega=\Lambda^{-1}d\Lambda$, the torsion constraint becomes $d(\Lambda \cE \Lambda^{-1})=0$, which is easily solved.}
\begin{align}
\Omega_i&=\Lambda^{-1}\d_i\Lambda,\label{eqw}\\ 
\cE_i&=\Lambda^{-1}\d_ia\Lambda.\label{eqe}
\end{align}
where $\Lambda$ is an element of $SL(3,\mathbb{R})$ and $a$ lives on its algebra. The idea now, is to use these fields in order to construct an action principle entirely defined at the boundary. However, the field content depends on variables located not only at the boundary, but also at the bulk ($r$ coordinate). In order to parameterize this $r$-dependence, we will first impose partial gauge-fixing conditions, which in this case will be given by
\begin{align}
\cE_r=f(r),\quad \Omega_r=0,
\end{align}
where $f$ is a Lie algebra valued function. It is important to note that these conditions are compatible with the on-shell solution \eqref{Aonshell} but they differ from the one used in \cite{Barnich:2013yka}. Also, they imply that $\Lambda$ and $a$ can be written as
\begin{align}
\Lambda&=\lambda(u,\phi),\label{eql}\\
a&=\alpha(u,\phi)+\lambda(u,\phi)\beta(r)\lambda^{-1}(u,\phi) \label{eqa},
\end{align}
where $\beta$ is defined through the relation $\d_r \beta = f$. Hence, these relations lead to modified expressions for the generalized spin connection and dreibein,
\begin{align}
\Omega_i&=\lambda^{-1}\d_i\lambda,\\
\cE_i&=\lambda^{-1}\d_i \alpha \lambda + \d_i \beta +[\lambda^{-1}\d_i \lambda, \beta ].
\end{align}
By replacing these expressions back in \eqref{WdAc},  we obtain the following action principle,
\begin{equation}
\label{WZWflat}
\tilde{I}[\lambda,\alpha]=\frac{k}{\pi}\int du d\phi\; {\rm Tr}\big[\dot{\lambda}\lambda^{-1}\alpha'-\frac{1}{2}(\lambda^{-1} \lambda')^2\big].
\end{equation}
Note that this model was first obtained in the $iso(2,1)$ case, when analog boundary conditions were taken into account \cite{Barnich:2013yka}. There, it was shown that  this theory can be obtained as a flat limit of the sum of two chiral WZW actions. For this reason, we will refer to \eqref{WZWflat} as ``Flat WZW model''. A similar action was also found in \cite{Salomonson:1989fw} for boundary conditions not related to asymptotically flat spacetimes at null infinity.

\section{Symmetries of Flat WZW action}
\label{symWZW}

In what follows we consider action \eqref{WZWflat} where $\lambda$ is an element of $SL(3,\mathbb{R})$ and $\alpha$ belongs to its Lie algebra. For simplicity, we will refer to the $J_a$ and $J_{ab}$ matrices simply as $T_A$, where capital latin indexes runs from $1$ to $8$. In this notation, the Killing metric and the structure constants can be computed as
\begin{align}
  \label{eq:tensores}
  g_{AB}=2{\rm Tr}(T_AT_B),\quad f_{ABC}=2{\rm Tr}([T_A,T_B]T_C).
\end{align}
Algebra indexes will be raised and lowered with $g^{AB}$ and $g_{AB}$ respectively. Note that, both $g_{AB}$ and $f_{ABC}$ can be extracted from relations \eqref{traza} and \eqref{sl3}.

Let us now review general features of this model. As with chiral WZW \cite{Moore:1989yh,Elitzur:1989nr}, this action contains a first class constraint which gives rise to gauge transformations 
\begin{equation}
\label{zeromodes}
\delta \lambda= \epsilon(u) \lambda \quad \delta \alpha= [\epsilon(u),\alpha].
\end{equation}
Besides from this time-dependent transformation, this action is invariant under two global symmetries. We will show that the symmetry algebra of the associated currents corresponds to an affine extension of \eqref{algs3}. 

The first transformation is given by,
\begin{eqnarray}
\delta\lambda&=&0\label{t1a},\\
\delta\alpha&=&\lambda\sigma\lambda^{-1}\label{t1b},
\end{eqnarray}
where  $\sigma$ is a Lie algebra valued function depending only on $\phi$.  By means of Noether theorem, the components of the conserved current $\partial_\mu \cP^\mu=0$ generated by this symmetry, are
\begin{subequations}
\label{cons1}
\begin{align}
\cP^u&= \sigma_{A}\cP^A, \quad \cP^A =\frac{k}{2\pi}[\lambda^{-1} \lambda']^A, \\ 
\cP^\phi&=0. 
\end{align}
\end{subequations}
The second symmetry correspond to
\begin{eqnarray}
\delta\lambda&=&-\lambda \theta, \label{t2a}\\
\delta\alpha&=&-u\lambda\theta' \lambda^{-1},\label{t2b}
\end{eqnarray}
the parameter $\theta$ is some Lie algebra element depending only on the angular coordinate. Its associated conserved current $\partial_\mu \cJ^\mu=0$ can be cast into
\begin{subequations}
\label{cons2}
\begin{align}
\cJ^u&= \theta_{A}\cJ^A, \quad \cJ^A=-\frac{k}{2\pi}[\lambda^{-1}\alpha'\lambda -u(\lambda^{-1}\lambda')']^A,\\ 
\cJ^\phi&=0. 
\end{align}
\end{subequations}
For the $iso(2,1)$ case, the algebra of these currents was worked out in \cite{Barnich:2013yka} after a lengthy Hamiltonian analysis. Here we will adopt the approach of \cite{Barnich:2013axa}, where it was shown that for a given current $J_X$, where $X$ stands for the parameter of the transformation, the Poisson bracket $\{J_{X_1},J_{X_2}\}$ can be read from,
\begin{equation}
\delta_{X_2} J_{X_1}= J_{[X_1,X_2]}+K_{X_1,X_2},\label{truco}
\end{equation}
where  $\delta_{X_2} J_{X_1}$ is the variation of $J_{X_1}$ with respect to the symmetry $X_2$, $[X_1,X_2]$ represents the Lie bracket and $K_{X_1,X_2}$ is a potentially non-vanishing central extension.
In order to apply this procedure, we will define the smeared currents,
\begin{equation}
\cP(\sigma)=\int d\phi \sigma^A \cP_A, \quad \cJ(\theta)=\int d\phi \theta^A \cJ_A.
\end{equation}
Then, by a direct application of \eqref{truco} we obtain,
\begin{eqnarray}
\{\cP(\theta),\cP(\sigma)\}&=&0,\\
\{\cJ(\theta),\cJ(\sigma)\}&=&\mathcal{J}([\theta,\sigma]),\\
\{\cJ(\theta),\cP(\sigma)\}&=&\mathcal{P}([\theta,\sigma])+\frac{k}{2\pi}\int d\phi \theta'^A \sigma_A,
\end{eqnarray}
where $[\theta,\sigma]^A=f^{A}_{\;\;\;BC}\theta^B \sigma^C $. If we write the Poisson brackets relations in terms of $\cJ_A$ and $\cP_A$, they become
\begin{subequations}
\label{currentalg}
\begin{eqnarray}
\label{currentJP}
\{\cP_A(\phi),\cP_B(\phi')\}&=&0,\\
\{\cJ_A(\phi),\cJ_B(\phi')\}&=&f_{AB}^{\;\;\;\;\:C}\mathcal{J}_C(\phi) \delta(\phi-\phi'),\\
\{\cJ_A(\phi),\cP_B(\phi')\}&=&f_{AB}^{\;\;\;\;\:C}\mathcal{P}_C(\phi)\delta(\phi-\phi')-\frac{k}{2\pi}g_{AB}\partial_\phi \delta(\phi-\phi'),
\end{eqnarray}
\end{subequations}
which is the affine extension of algebra \eqref{algs3}. In the next section we will show that the higher spin extended $bms_3$ algebra can be obtained from this algebra through a generalization of Sugawara construction including higher spin charges.

\section{Extended Sugawara Construction}
\label{sec:2ndbms}

The chiral WZW flat model \eqref{WZWflat}  based on $sl(3,\mathbb{R})$ Lie algebra contains
a generalization of the $bms_3$ algebra involving generators with spin $s=2,3$.  Indeed,
the spin-two densities are constructed as quadratic invariants of the currents,
\begin{align}\label{MJ}
\cM=\frac{\pi}{k}g_{AB}\cP^A\cP^B, \quad \cJ=-\frac{2\pi}{k}g_{AB}\cP^A\cJ^B.
\end{align}
Using \eqref{currentalg}, the algebra of these densities gives
\begin{equation}
  \label{eq:1}
  \begin{split}
 \{\cM(\phi),\cM(\phi')\}&=0,\\
\{\cM(\phi),\cJ(\phi')\}&=(\cM(\phi)+\cM(\phi'))\d_\phi\delta(\phi-\phi'),\\
\{\cJ(\phi),\cJ(\phi')\}&=(\cJ(\phi)+\cJ(\phi'))\d_\phi\delta(\phi-\phi'),
  \end{split}
\end{equation}
which is recognized as $bms_3$ algebra without central extensions. Unlike the $sl(2,\mathbb{R})$ case, the greater rank algebra which we are working with allows to define cubic invariants\footnote{One could also consider quantities such as $$g_{AB}\cJ^A\cJ^B,\quad d_{ABC}\cJ^A\cJ^B\cJ^C,\quad \text{or} \quad d_{ABC}\cJ^A\cJ^B\cP^C,$$ but we will not use them since they vanish when \eqref{ii} and \eqref{iii} are implemented.}
\begin{align}
\label{dens3}
\quad \cW=\frac{4\pi^2}{3k^2}d_{ABC}\cP^A\cP^B\cP^C, \quad \cV=-\frac{4\pi^2}{k^2}d_{ABC}\cP^A\cP^B\cJ^C,
\end{align}
where $d_{ABC}$ is an invariant $sl(3,\mathbb{R})$ tensor constructed as
\begin{equation}
  \label{eq:ds}
  d_{ABC}=2{\rm Tr}([T_A,T_B]_{+}T_C),
\end{equation}
with $[ \;,\; ]_+$ representing the anti-commutator. Note that, by definition,  this tensor is completely symmetric. 

The Poisson brackets between the quadratic and the cubic generator read,
\begin{equation}
  \label{eq:2}
  \begin{split}
\{\cM(\phi),\cW(\phi')\}&=0,\\
\{\cJ(\phi),\cW(\phi')\}&=(\cW(\phi)+2\cW(\phi'))\d_\phi\delta(\phi-\phi'),\\
\{\cM(\phi),\cV(\phi')\}&=(\cW(\phi)+2\cW(\phi'))\d_\phi\delta(\phi-\phi'),\\
\{\cJ(\phi),\cV(\phi')\}&=(\cV(\phi)+2\cV(\phi'))\d_\phi\delta(\phi-\phi'),
  \end{split}
\end{equation}
and finally, by using relation \eqref{dd},  we obtain
\begin{equation}
  \label{eq:3}
  \begin{split}
\{\cW(\phi),\cW(\phi')\}&=0,\\
\{\cW(\phi),\cV(\phi')\}&=\frac{32 \pi}{3k}(\cM(\phi)^2+\cM(\phi')^2)\d_\phi\delta(\phi-\phi'),\\
\{\cV(\phi),\cV(\phi')\}&=\frac{16\pi}{3k}(\cM(\phi)\cJ(\phi)+\cM(\phi')\cJ(\phi'))\d_\phi\delta(\phi-\phi').
  \end{split}
\end{equation}
Algebra \eqref{eq:1}, \eqref{eq:2}, \eqref{eq:3} can be regarded as a spin-3 extension of $bms_3$. The main difference of this algebra with the one found in \cite{Afshar:2013vka,Gonzalez:2013oaa} is the central term.
In order to incorporate central extensions we have to take into account boundary conditions \eqref{ii}. This requires
the introduction of gauge fixing conditions \eqref{iii} in such a way that the complete system of reduction conditions become second class. This will be discussed in section \bref{B}.

\section{Reduced flat WZW model}
\label{RA}
In this section, we implement the remaining conditions in the flat WZW model. The purpose is twofold: first, in section \bref{A}, a reduced theory will be obtained when solving the condition \eqref{ii} inside of the action. On the other hand, in section \bref{B}, conditions \eqref{ii}, \eqref{iii} will be used to implement Dirac brackets of densities \eqref{MJ}, \eqref{dens3} defined in the previous section. 

Both conditions \eqref{ii} and \eqref{iii} can be translated into a fixation of some components of the conserved currents \eqref{cons1} and \eqref{cons2}. Indeed, conditions \eqref{ii} can be written as\footnote{Recall that capital latin indexes runs from $1$ to $8$, such that $$\{T_1,T_2,T_3,T_4,T_5,T_6,T_7,T_8\}=\{J_{0},J_{1},J_{2},J_{00},J_{01},J_{02},J_{11},J_{12}\}.$$}
\begin{align}
\label{1class}
  \psi_1=\cP^2-\frac{k}{2\pi}, \quad \psi_2=\cJ^2, \quad \psi_3=\cP^7,\quad \psi_4=\cJ^7,\quad \psi_5=\cP^8,\quad \psi_6 =\cJ^8.
\end{align}
Under current algebra \eqref{currentalg},  this set is shown to be first class. Furthermore, conditions \eqref{iii} in terms of the current components become
\begin{align}
\label{2class}
 \chi_1=\cP^3, \quad \chi_2=\cJ^3, \quad \chi_3=\cP^5, \quad \chi_4=\cJ^5, \quad \chi_5=\cP^6, \quad \chi_6=\cJ^6.
\end{align}

\subsection{Reduction of the action}
\label{A}
We will perform the reduction at the level of the action \eqref{WZWflat}. The outcome of this procedure will be a theory possessing the higher-spin-extended $bms_3$ as a global symmetry.

To accomplish this task, it is better to decompose the group element $\lambda$ as,  
\begin{equation}\label{gauss}
\begin{split}
&\lambda=ABC,\\
A=\exp\left(\sum_{a} \sigma^a E^{-}_a \right) \quad B=&\exp\left(-\frac{1}{2}\sum_{i} \varphi^{i} H_{i} \right), \quad C=\exp\left(\sum_{a} \tau^{a} E^{+}_a \right),
\end{split}
\end{equation}
where we have made use of the Gauss decomposition, using the Cartan basis for $sl(3,\mathbb{R})$. This representation consists of two commuting generators $H_i$ and six generators $E^{\pm}_a$ labeled by roots $a$ (See appendix \bref{a1}  for details on conventions and the relation with basis \eqref{sl3}). Fields $\sigma^a$, $\varphi^i$, $\tau^a$  are arbitrary scalars. 

The basis $H_i,E_a^{\pm}$ can always be chosen to fulfill the following trace relations
\begin{equation}
\label{traces}
\text{Tr}[H_iH_j]=K_{ij}, \quad \text{Tr}[E_a^{\pm} E_b^{\mp}]=\delta_{ab},\quad  \text{Tr}[E_a^{\pm} E_b^{\pm}]=0, \quad  \text{Tr}[E_a^{\pm} H_i]=0,
\end{equation}
where $K_{ij}$ corresponds to the Cartan matrix of $sl(3,\mathbb{R})$. Also, it will be of particular importance to identify the generators $E^{\pm}_a$ labeled by simple roots, which will be denoted as\footnote{Note that $a_i$ refers to the $i$-th root, not to the $i$-th component of the root $a$.},
\begin{align}
E_{a_i}^{\pm}\equiv E_{i}^{\pm} \quad \text{when $a_i$ is a simple root}.  
\end{align}
Note that there are as many simple roots as generators $H_{i}$.

Reduction conditions \eqref{1class} can be translated in terms of the Cartan basis. First, those which involve components of $\mathcal{P}$, i.e., $\psi_1$, $\psi_3$ and $\psi_5$, can be expressed as 
\begin{equation}
\label{consP1}
(\lambda^{-1} \lambda')^{(-)} = \sum_{i} \mu^{i} E^{-}_{i},
\end{equation}
where the superscript $(-)$ denotes the components along the Lie algebra element $E^{-}_{a}$ and the sum is taken over the negative simple roots. The coefficients $\mu^i$ are given by
$$\mu^{1}=\mu^{2}=\frac{1}{\sqrt{2}}.$$

It is convenient to rewrite condition \eqref{consP1} in terms of the Gauss decomposition of $\lambda$. This can be accomplished by making use of the trace relations \eqref{traces} and the commutation relations \eqref{cheva1}, \eqref{cheva2} given in appendix \bref{a1}. Using this, we find
\begin{equation}
\label{consP11}
A^{-1} A' = \sum_{i} \mu^{i} e^{\frac{1}{2}\sum_j K_{ij}\varphi^j } E^{-}_{i}.
\end{equation}

On the other hand, reductions conditions involving components of $\cJ$, i.e., $\psi_2$, $\psi_4$ and $\psi_6$, imply 
\begin{equation}
\label{consJ2}
(\lambda^{-1} \alpha' \lambda)^{(-)} = 0,
\end{equation}
which, in terms of Gauss decomposition, can be simplified to
\begin{equation}
\label{consJ22}
(B^{-1}A^{-1} \alpha' A B)^{(-)} = 0.
\end{equation}

With reduction conditions \eqref{consP11} and \eqref{consJ22} at hand, we are ready to apply them inside of the action. The flat WZW action \eqref{WZWflat} can be expressed as
\begin{multline}
\tilde{I}[ABC,\alpha]=\frac{k}{\pi}\int du d\phi \  {\rm Tr}\left[-(A^{-1}A')^{\centerdot} A^{-1}\alpha A+\dot{B}B^{-1}A^{-1}\alpha'A+ \dot{C}C^{-1}(B^{-1}A^{-1}\alpha' A B )\right. \\
\left.
-\half (A^{-1}A')^2-\half (B^{-1}B')^2-\half (C^{-1}C')^2- A^{-1}A' B'B^{-1}-B^{-1}B'C'C^{-1}-(B^{-1}A^{-1}A' B) C'C^{-1}
\right],
\end{multline}
where we have discarded boundary terms in order to impose directly relations \eqref{consP11} and \eqref{consJ22}. In fact, by using relations \eqref{traces}, one obtains
\begin{align}
\label{Ired}
\tilde{I}_R[\xi_{i},\varphi^{i}]=\frac{k}{4\pi}\int du d\phi \  \left(\sum_{i}(\xi_{i})'  \dot{\varphi}^{i}-\frac{1}{2}\sum_{i,j}K_{i j} (\varphi^i)^{\prime} (\varphi^j)^{\prime} \right),
\end{align}
where $\xi_i=-2{\rm Tr}[H_i A^{-1} \alpha A]$. Action \eqref{Ired} is the natural generalization to $sl(3,\mathbb{R})$ of the results found in \cite{Barnich:2013yka}. 

Note that, apart from the global symmetry transformations, this model also inherits the zero mode transformations \eqref{zeromodes} present in the flat WZW action
\begin{equation}
\label{zeromodes2}
\delta \varphi^{i}=f^{i}(u), \quad \delta \xi^{i}=g^{i}(u),
\end{equation}
where $f^i$ and $g^i$ are arbitrary functions. In section \bref{Toda}, it will be shown that action \eqref{Ired} is related to a suitable limit of $sl(3,\mathbb{R})$ Toda theory.
However, this latter theory is devoid of transformations \eqref{zeromodes2}.

\subsection{Dirac charge algebra}
\label{B}
The next step is to determine the Poisson structure associated to the charges of the reduced action \eqref{Ired}.  To do so, we must implement first class conditions \eqref{1class} inside of algebra \eqref{eq:1}, \eqref{eq:2}, \eqref{eq:3}. This is achieved by imposing gauge fixing conditions \eqref{2class} and constructing Dirac brackets.  

If we denote as $\Phi=\{\psi,\chi\}$ the set of all constraints, the Dirac bracket between any quantities $A$ and $B$ read,
\begin{equation}
\label{dB}
  \{A(\phi),B(\phi')\}^{*}=\{A(\phi),B(\phi')\}-\int d\phi_1 d\phi_2 \{A(\phi),\Phi_\alpha(\phi_1)\} \left[C^{\alpha\beta}(\phi_1,\phi_2)\right]^{-1}\{\Phi_\beta(\phi_2),B(\phi')\},
\end{equation}
where $C_{\alpha \beta}(\phi,\phi')=\{\Phi_\alpha(\phi),\Phi_\beta(\phi')\}$ is the constraint Poisson bracket matrix. 
By a straightforward, albeit tedious computation, one is able to find the bracket of densities $\cM$, $\cJ$, $\cW$, $\cV$ on the surface defined by constraints $\Phi$, leading to
\begin{equation}
\label{algD1}
\begin{split}
  \{\cM(\phi),\cM(\phi')\}^{*}=&0,\\
  \{\cM(\phi),\cJ(\phi')\}^{*}=&(\cM(\phi)+\cM(\phi'))\d_\phi\delta(\phi-\phi')-\frac{k}{2\pi}\d_\phi^3\delta(\phi-\phi'),\\
  \{\cJ(\phi),\cJ(\phi')\}^{*}=&(\cJ(\phi)+\cJ(\phi'))\d_\phi\delta(\phi-\phi').
  \end{split}
\end{equation}
This is the standard $bms_3$ algebra of pure General Relativity in three dimensions \cite{Barnich:2006av}. Furthermore, generators $\cW$ and $\cV$ still transform with spin-three
\begin{equation}
\label{algD2}
\begin{split}  
  \{\cM(\phi),\cW(\phi')\}^{*}=&0,\\
  \{\cM(\phi),\cV(\phi')\}^{*}=&(\cW(\phi)+2\cW(\phi'))\d_\phi\delta(\phi-\phi'), \\
  \{\cJ(\phi),\cW(\phi')\}^{*}=&(\cW(\phi)+2\cW(\phi'))\d_\phi\delta(\phi-\phi'),\\
  \{\cJ(\phi),\cV(\phi')\}^{*}=&(\cV(\phi)+2\cV(\phi'))\d_\phi\delta(\phi-\phi'),\\
  \end{split}
\end{equation}
however, the algebra among the spin-three generators get modified by linear and central terms,
\begin{equation}
\label{algD3}
\begin{split}  
 \{\cW(\phi),\cW(\phi')\}^{*}=&0,\\
 \{\cW(\phi),\cV(\phi')\}^{*}=&\frac{16 \pi}{3k}(\cM(\phi)^2+\cM(\phi')^2)\d_\phi\delta(\phi-\phi')\\
  &-\frac{1}{3}\big[2\d_\phi^3 \cM(\phi)\delta(\phi-\phi')  +9\partial_\phi^2\cM(\phi)\d_\phi\delta(\phi-\phi')\\
  &+15\partial_\phi \cM(\phi)\partial_\phi^2\delta(\phi-\phi') +10\cM(\phi)\partial_\phi^3\delta(\phi-\phi') -\frac{k}{2 \pi}\partial_\phi^5\delta(\phi-\phi') \big],\\
  \{\cV(\phi),\cV(\phi')\}^{*}=&\frac{32 \pi}{3k}(\cM(\phi)\cJ(\phi)+\cM(\phi')\cJ(\phi'))\partial_\phi\delta(\phi-\phi')\\
  &-\frac{1}{3}\big[2\partial_\phi^3 \cJ(\phi)\delta(\phi-\phi')  +9\partial_\phi^2\cJ(\phi)\partial_\phi\delta(\phi-\phi')\\
  &+15\partial_\phi \cJ(\phi)\partial_\phi^2\delta(\phi-\phi') +10\cJ(\phi)\partial_\phi^3\delta(\phi-\phi')\big].
  \end{split}
\end{equation}
When expressing in Fourier modes, one obtains the same result as in \cite{Gonzalez:2013oaa}. Indeed, definitions 
\begin{align}
\label{modes}
P_n=\int^{2\pi}_{0}d\phi \ e^{in\phi} \cM, \quad J_n=\int^{2\pi}_{0}d\phi \ e^{in\phi} \cJ, \quad W_n=\int^{2\pi}_{0} d\phi \ e^{in\phi} \cW, \quad V_n=\int^{2\pi}_{0} d\phi \ e^{in\phi} \cV, 
\end{align}
lead to the higher spin extension of $bms_3$ with central charge
\begin{equation}
c=12k=\frac{3}{G}.
\end{equation}
The explicit form of the constraint matrix, some intermediate expressions needed to evaluate \eqref{dB} and the above algebra expressed in modes will be presented in appendix \bref{a2}. 


\section{Connection with flat limit of Toda theories}
\label{Toda}
We will prove that action \eqref{Ired} can be regarded as a suitable flat limit of $sl(3,\mathbb{R})$ Toda theory. In order to demostrate this assertion,  let us specify an appropriate limit.
  
In its Hamiltonian form, Toda theory on a cylinder of radius $l$ is defined by,
\begin{align}
\label{TodaH}
I_{\text{Toda}}[\pi_i,\varphi^{i}]=\int du d\phi \left( \sum_{i} \pi_{i} \dot{\varphi}^{i} -\sum_{i,j} \frac{1}{2}\left[G^{ij}\pi_i \pi_j+ \frac{1}{l^2}K_{ij}(\varphi^i)^{\prime} (\varphi^j)^{\prime}\right]-\sum_{i}M^{i}e^{\frac{1}{2} \gamma \sum_{j} K_{i j} \varphi^{j}} \right),
\end{align}
where $G^{ij}$ is the inverse of the Cartan matrix, fields $\varphi^i$ and $\pi_{i}$ are canonical pairs, whereas $\gamma$ and $M^{i}$ are coupling constants. This theory is known to be invariant under two copies of $W$-symmetry \cite{Bilal:1988jf,Bilal:1988ze,Balog:1990mu}.

The flat limit of \eqref{TodaH} consists in taking $l \to \infty$. However, as it occurs with Liouville theory \cite{Barnich:2012rz}, by simply taking the limit, the resulting theory possesses a charge algebra which does not include central terms, hindering the connection with asymptotically flat spacetimes in three dimensions. Therefore, along the lines of \cite{Barnich:2012rz}, we first rescale the fields as
\begin{equation}
\pi_i=\frac{\Pi_i}{l} \quad \varphi^{i}=l\Phi^{i},
\end{equation}
and then take $l\to \infty$, keeping $M^{i}$ and $\beta=\gamma l$ fixed, obtaining
\begin{align}
\label{TodaHP}
I_\text{flat-Toda}[\Pi_i,\Phi^{i}]=\int du d\phi \left( \sum_{i} \Pi_{i} \dot{\Phi}^{i} -\sum_{i,j} \frac{1}{2}K_{ij}(\Phi^i)^{\prime} (\Phi^j)^{\prime}-\sum_{i}M^{i}e^{\frac{1}{2} \beta \sum_{j} K_{i j} \Phi^{j}} \right).
\end{align}
It is important to stress that this is a first order action which does not have a second order counterpart. 

The interesting observation is that flat-Toda action can be related to \eqref{Ired} by a field redefinition. In fact, by defining
\begin{equation}
\label{trans}
\begin{split}
\Pi_i &=\frac{\sqrt{2}}{\beta} (\xi_{i})^{\prime}-\frac{\beta u}{2}\sum_{k} M^{k}K_{ki} e^{\frac{1}{\sqrt{2}} \sum_{j} K_{k j} \varphi^{j}} ,\\
\Phi^{i}&=\frac{\sqrt{2}}{\beta}\varphi^i,
\end{split}
\end{equation}
and replacing back in \eqref{TodaHP}, we obtain action \eqref{Ired} (up to a boundary term), provided $\beta^2=\frac{8\pi}{k}$. 

A few comments on \eqref{trans} are in order. A similar kind of transformation has been previously used to obtain the $bms_3$ invariant Liouville theory from the corresponding model \eqref{Ired} \cite{Barnich:2013yka}. 
Nonetheless, transformation \eqref{trans} differs since it is explicitly time dependent. However, in both cases,  zero modes \eqref{zeromodes2} are dropped by these field redefinitions. The origin of this exclusion can be traced back to the identification of $\Pi_i$ with $(\xi_{i})^{\prime}$, which is not invertible in the zero mode sector. This feature has been previously observed when two chiral bosons are combined giving rise to Liouville theory \cite{Henneaux:1999ib}.   

\section{Comments on higher rank groups}
\label{higher}

Results presented in the previous sections are easily generalized beyond the $sl(3,\mathbb{R})$ case. If we consider the model \eqref{WZWflat} for $\lambda \in SL(n,\mathbb{R})$ and $\alpha \in sl(n,\mathbb{R})$, we can construct densities with spin $s$, where $3 \le s \le n$. Indeed, they can be written as
\begin{equation}
\cW_{s}=\frac{1}{s}\left(\frac{2\pi}{k}\right)^s d_{A_1 \cdots A_s} \cP^{A_1} \cdots \cP^{A_s}, \quad \cV_{s}=-\left(\frac{2\pi}{k}\right)^s d_{A_1 \cdots A_s} \cP^{A_1} \cdots \cP^{A_{s-1}}\cJ^{A_s},
\end{equation}
where $d_{A_1 \cdots A_s}$ are symmetric invariant tensors of $sl(n,\mathbb{R})$. The commutation relations of these quantities with $\cM$ and $\cJ$ become
\begin{equation}
  \label{eq:2s}
  \begin{split}
\{\cM(\phi),\cW_{s}(\phi')\}&=0,\\
\{\cJ(\phi),\cW_{s}(\phi')\}&=\left[\cW_{s}(\phi)+(s-1)\cW_{s}(\phi')\right]\d_\phi\delta(\phi-\phi'),\\
\{\cM(\phi),\cV_{s}(\phi')\}&=\left[\cW_{s}(\phi)+(s-1)\cW_{s}(\phi')\right]\d_\phi\delta(\phi-\phi'),\\
\{\cJ(\phi),\cV_{s}(\phi')\}&=\left[\cV_{s}(\phi)+(s-1)\cV_{s}(\phi')\right]\d_\phi\delta(\phi-\phi').
  \end{split}
\end{equation}
The above expressions show that $\cV_{s}$ and $\cW_{s}$ transform as spin-$s$ generators. Hence, we expect that the whole set $\{\cM,\cJ,\cW_{3},\cV_{3}....\cW_{n},\cV_{n}\}$ spans the spin-$n$ extension of $bms_3$ algebra.  

Note that for the $sl(n,\mathbb{R})$ case, the asymptotic form of the connection $A$ was given in \cite{Gonzalez:2013oaa}

\begin{align}
A=&\left( \frac{1}{2}\mathcal{M}du-dr+\left( \mathcal{J}+\frac{u}{2}\d_{\phi }\mathcal{M}\right) d\phi \right)P_{0}+du P_{1}+rd\phi P_{2}+\frac
{1}{2}\mathcal{M}d\phi J_{0}+d\phi J_{1}\nonumber\\
& +\left(  \mathcal{W}_{3}du+\left(  \mathcal{V}_{3}+u\partial_{\phi
}\mathcal{W}_{3}\right)  d\phi\right)  P_{00}+\mathcal{W}_{3}d\phi
J_{00}\label{A flat reloaded}\\
& +\left(  \mathcal{W}_{4}du+\left(  \mathcal{V}_{4}+u\partial_{\phi
}\mathcal{W}_{4}\right)  d\phi\right)  P_{000}+\mathcal{W}_{4}d\phi
J_{000}\nonumber\\
& +...\nonumber
\end{align}

These generalized boundary conditions does not alter significantly the form of the reduction relations \eqref{ii} and \eqref{iii}, leaving \eqref{i} unchanged. As a result, the procedure carried out in section \bref{RA} is still true for higher order groups. In this case, \eqref{ii} is translated into conditions on the components of the currents along the negative simple roots of the Cartan basis. Accordingly, the reduced model acquires the same form as in \eqref{Ired} provided $K_{ij}$ being the Cartan matrix of $sl(n,\mathbb{R})$.

 The full spin-$n$ extension of $bms_3$ algebra, along with the Dirac bracket computation, will be presented elsewhere.
 
 \section{Conclusions}
 
We have built the two-dimensional action principle invariant under a spin-three extension of BMS$_3$ group, which at classical level corresponds to the dual of asymptotically flat gravity coupled to spin-three fields. The charges associated to this theory span an algebra which coincides with the previous analysis \cite{Gonzalez:2013oaa,Afshar:2013vka}.  We have shown that this latter theory is related to a flat spacetime limit of Toda theory. 

The importance of these results is twofold. From the point of view of holography, these reduced models are appropriate candidates to understand a quantum duality of higher spin gravity on asymptotically flat geometries. On the other hand, the properties of these nonlinear algebras can be better understood, in particular, in its relation with $W$-algebras. In this sense, this work supports the fact that asymptotically flat spacetimes have structures as rich as spaces with AdS$_3$ asymptotics. 

\acknowledgments We thank G. Barnich, M. Ba\~nados, A. Campoleoni, R. Canto, J. Gamboa, J. Matulich, P. Salgado-Rebolledo,  C. Troessaert and R. Troncoso for useful discussions and comments. H.G. is supported in part by IISN-Belgium, and by ``Communaut\'e fran\c{c}aise de Belgique - Actions de Recherche Concert\'ees''. M.P. is partially supported by Fondecyt (Chile) $\#$11130083 and $\#$7912010045. M.P. also thanks ``Service de Physique Math\'ematique des Interactions Fondamentales" at Universit\'e Libre de Bruxelles, where part of this work was done.

\appendix

\section{Conventions}
\label{a1}

Throughout this work, the following conventions on Lie algebras were used. 

The $sl(3,\mathbb{R})$ invariants tensors are defined as
\begin{eqnarray}
g_{AB}&=&2{\rm Tr}[T_A, T_B],\\
f_{ABC}&=&2{\rm Tr}[[T_A, T_B]T_C],\\
d_{ABC}&=&2{\rm Tr}[[T_A, T_B]_{+}T_C].
\end{eqnarray}

From the above definitions, $f_{ABC}$ and $d_{ABC}$ must fulfill Jacobi identities 
\begin{eqnarray}
f_{BIA}f^A_{\:\:CE}+f_{EIA}f^A_{\:\:BC}+f_{CIA}f^A_{\:\:EB}=0,\\
f_{BIA}d^A_{\:\:CE}+f_{EIA}d^A_{\:\:BC}+f_{CIA}d^A_{\:\:EB}=0.
\end{eqnarray}

A relation among the above quantities, used in section \bref{sec:2ndbms}
\begin{eqnarray}
\label{dd}
g^{CH} d_{ABC} d_{HDE}=\frac{4}{3}\left(
 g_{AD} g_{BE}+ g_{AD} g_{BE}- g_{AB} g _{DE}\right)
-\frac{1}{3}g^{CH}( f_{ADC} f_{HDE}+ f_{AEC} f_{HBD}).
\end{eqnarray}

We have used two different basis for $sl(3,\mathbb{R})$. First, the $J_a,J_{ab}$ basis, fulfilling commutations relations \eqref{sl3}, corresponds to  
\begin{equation}
\footnotesize
\begin{split}
J_0=
  \begin{pmatrix}
    0 &  \sqrt{2} & 0 \\
    0 & 0 & \sqrt{2} \\
    0 & 0 & 0 
  \end{pmatrix},
  \quad
  J_1&=
  \begin{pmatrix}
    0 &  0 & 0 \\
    \frac{1}{\sqrt{2}} & 0 & 0 \\
    0 &  \frac{1}{\sqrt{2}}  & 0 
  \end{pmatrix},
  \ \
  J_2=
  \begin{pmatrix}
    1 &  0 & 0 \\
    0 & 0 & 0 \\
    0 & 0 & -1 
  \end{pmatrix},\\
  J_{00}=
  \begin{pmatrix}
    0 &  0 & -4 \\
    0 & 0 & 0 \\
    0 & 0 & 0 
  \end{pmatrix},
    \ \
  J_{01}=
  \begin{pmatrix}
    \frac{1}{3} &  0 & 0 \\
    0 & -\frac{2}{3} & 0 \\
    0 & 0 & \frac{1}{3} 
  \end{pmatrix},
  \ \
  J_{02}&=
  \begin{pmatrix}
    0 &  -\sqrt{2} & 0 \\
    0 & 0 & \sqrt{2} \\
    0 & 0 & 0 
  \end{pmatrix},
    \ \
  J_{11}=
  \begin{pmatrix}
    0 &  0 & 0 \\
    0 & 0 & 0 \\
    -1 & 0 & 0 
  \end{pmatrix},
  \ \
  J_{12}=
  \begin{pmatrix}
    0 &  0 & 0 \\
    -\frac{1}{\sqrt{2}} & 0 & 0 \\
    0& \frac{1}{\sqrt{2}} & 0 
  \end{pmatrix}.
  \end{split}
\end{equation}
We assume a non-diagonal Minkowski metric in tangent space, whose only non vanishing components are given by $\eta_{01}=\eta_{10}= \eta_{22} = 1$, and the Levi-Civita symbol fulfills $\epsilon_{012} = 1$.

In section \bref{RA}, we have made use of Chevalley basis. The algebra spanned by generators $H_i$ and $E^{\pm}_i$ takes the form 
\begin{equation}\label{cheva1}
[H_i,H_j]=0,\quad [H_i,E^{\pm}_j]=\pm K_{ji}E^{\pm}_j, \quad [E^+_i,E^-_j]=\delta_{ij}H_j.
\end{equation}
The rest of the algebra can be obtained by repeated commutations of $E^{\pm}_i$. It gives 
\begin{equation}\label{cheva2}
\begin{split}
[E^{\pm}_i,E^{\pm}_3]&=0,\quad [E^{+}_3,E^{-}_3]=H_{1}+H_{2}, \quad [H_i,E^\pm_3]=\pm E^{\pm}_3,\\
[E^\pm_1,E^\pm_2]&=\pm E^{\pm}_3, \quad [E^\pm_1,E^\mp_3]=\mp E^{\mp}_2, \quad [E^\pm_2,E^\mp_3]=\mp E^{\mp}_1.
\end{split}
\end{equation}
The explicit matrices are given by,
\begin{equation}
\footnotesize
\begin{split}
H_1=
  \begin{pmatrix}
    1 &  0 & 0 \\
    0 & -1 & 0 \\
    0 & 0 & 0 
  \end{pmatrix},&
  \ \
  H_2=
  \begin{pmatrix}
    0 &  0 & 0 \\
    0 & 1 & 0 \\
    0 & 0 & -1 
  \end{pmatrix},\\
  E^{+}_1=
  \begin{pmatrix}
    0 &  1 & 0 \\
    0 & 0 & 0 \\
    0 & 0 & 0 
  \end{pmatrix},
  \ \
  E^{+}_2=
  \begin{pmatrix}
    0 &  0 & 0 \\
    0 & 0 & 1 \\
    0 & 0 & 0 
  \end{pmatrix},
  \ \
  E^{+}_3=
  \begin{pmatrix}
    0 &  0 & 1 \\
    0 & 0 & 0 \\
    0 & 0 & 0 
  \end{pmatrix},&
  \ \
  E^{-}_1=
  \begin{pmatrix}
    0 &  0 & 0 \\
    1 & 0 & 0 \\
    0 & 0 & 0 
  \end{pmatrix},
   \ \
  E^{-}_2=
  \begin{pmatrix}
    0 &  0 & 0 \\
    0 & 0 & 0 \\
    0 & 1 & 0 
  \end{pmatrix},
  \ \
  E^{-}_3=
  \begin{pmatrix}
    0 &  0 & 0 \\
    0 & 0 & 0 \\
    1& 0 & 0 
  \end{pmatrix}.
\end{split}
\end{equation}

The relation between the Chevalley basis and the $J_a,J_{ab}$ reads
\begin{equation}
\begin{split}
  H_1=\frac{1}{2}(J_2+3J_{01}), &\quad  H_2=\frac{1}{2}(J_2-3J_{01}),\\
E_1^{+}=\frac{\sqrt{2}}{4}(J_0-J_{02}),\quad E_2^{+}&=\frac{\sqrt{2}}{4}(J_0+J_{02}),\quad E_3^{+}=-\frac{1}{4} J_{00},\\
E_1^{-}=\frac{\sqrt{2}}{2}(J_1-J_{12}),\quad E_2^{-}&=\frac{\sqrt{2}}{2}(J_1+J_{12}),\quad E_3^{-}=- J_{11}.
\end{split}
\end{equation}

\section{Dirac Bracket computation}
\label{a2}
Here we display useful formulae for the computation of Dirac brackets. 
The Poisson brackets matrix of the constraints \eqref{1class} and \eqref{2class}$,  C_{\alpha\beta}(\phi,\phi')=\{\Phi_\alpha(\phi), \Phi_{\beta}(\phi')\}$, turns out be 
\begin{multline}
\nonumber
C_{\alpha \beta}=\\
  \bordermatrix{
    ~ & \psi_1 & \psi_2 &  \psi_3 & \psi_4 & \psi_5 & \psi_6 & \chi_1 & \chi_2 & \chi_3 & \chi_4 & \chi_5 & \chi_6 \cr
    \psi_1 & 0 & 0 &  0 & 0 & 0 & 0 & 0 & -\frac{k}{2\pi} \delta & 0 & 0 & 0 & 0 \cr
    \psi_2 & 0 & 0 &  0 & 0 & 0 & 0 & -\frac{k}{2\pi} \delta & 0 & 0 & 0 & 0 & 0 \cr
    \psi_3 & 0 & 0 &  0 & 0 & 0 & 0 & 0 & 0 & 0 & 0 & 0 & -\frac{k}{2\pi} \delta\cr
    \psi_4 & 0 & 0 &  0 & 0 & 0 & 0 & 0 & 0 & 0 & 0 &-\frac{k}{2\pi} \delta & 0\cr
    \psi_5 & 0 & 0 &  0 & 0 & 0 & 0 & 0 & 0 & 0 &\frac{3k}{2\pi} \delta & 0 &-\frac{k}{2\pi}\delta' \cr
    \psi_6 & 0 & 0 &  0 & 0 & 0 & 0 & 0 & 0 & \frac{3k}{2\pi} \delta & 0 & -\frac{k}{2\pi}\delta' & 0\cr
    \chi_1 & 0 & \frac{k}{2\pi} \delta &  0 & 0 & 0 & 0 & 0 & -\frac{k}{2\pi}\delta' & 0 & 0 & 0 & 0\cr
    \chi_2 & \frac{k}{2\pi} \delta & 0 &  0 & 0 & 0 & 0 &  -\frac{k}{2\pi}\delta' & 0 & 0 & 0 & 0 & 0\cr
    \chi_3 & 0 & 0 &  0 & 0 & 0 & -\frac{3k}{2\pi} \delta & 0 & 0 & 0 & -\frac{3k}{2\pi}\delta' & 0 & 3\cP^1 \delta\cr
    \chi_4 & 0 & 0 &  0 & 0 & -\frac{3k}{2\pi}\delta & 0 & 0 & 0 & -\frac{3k}{2\pi}\delta' & 0 & 3\cP^1 \delta & 3\cJ^1 \delta \cr
    \chi_5 & 0 & 0 &  0 & \frac{k}{2\pi} \delta & 0 & -\frac{k}{2\pi}\delta' & 0 & 0 & 0 & -3\cP^1 \delta & 0 & 0\cr
    \chi_6 & 0 & 0 &  \frac{k}{2\pi} \delta & 0 & -\frac{k}{2\pi}\delta' & 0 & 0 & 0 & -3\cP^1 \delta & -3\cJ^1 \delta & 0 & 0\cr
  }
\end{multline}
where $\delta\equiv\delta(\phi-\phi')$ and $\delta' \equiv \d_\phi\delta(\phi-\phi')$. We look for the inverse of $C_{\alpha\beta}$, in such a way that
\begin{equation}
  \int^{2\pi}_0 d\phi'' C^{-1\;\alpha\gamma}(\phi,\phi'')C_{\gamma\beta}(\phi'',\phi')=\delta^\alpha_\beta \delta(\phi-\phi'),
\end{equation}
where
\begin{multline}
\nonumber
\left[C^{\alpha\beta}\right]^{-1}=\\
  \begin{pmatrix}
    0 & -\frac{2\pi}{k}\delta' & 0 & 0 & 0 & 0 & 0 & \frac{2\pi}{k}\delta & 0 & 0 & 0 & 0 \\
    -\frac{2\pi}{k} \delta' & 0 & 0 & 0 & 0 & 0 &\frac{2\pi}{k}\delta & 0 & 0 & 0 & 0 & 0 \\
    0 & 0 & [C^{33}]^{-1} & [C^{34}]^{-1}& \frac{4\pi^2}{k^2}\cJ^1\delta & [C^{36}]^{-1} & 0 & 0 & 0 & -\frac{2\pi}{3k}\delta' & 0 & \frac{2\pi}{k}\delta \\
    0 & 0 & [C^{43}]^{-1} & 0 & [C^{45}]^{-1} & 0 & 0 & 0 & -\frac{2\pi}{3k}\delta' & 0 &  \frac{2\pi}{k}\delta & 0 \\
    0 & 0 & -\frac{4\pi^2}{k^2}\cJ^1 \delta & [C^{54}]^{-1}& 0 & -\frac{2\pi}{3k} \delta' & 0 & 0 & 0 &-\frac{2\pi}{3k}\delta  & 0 & 0 \\
    0 & 0  & [C^{63}]^{-1} & 0 &  -\frac{2\pi}{3k} \delta' & 0 & 0 & 0 &-\frac{2\pi}{3k}\delta  & 0 & 0 & 0 \\
    0 & -\frac{2\pi}{k}\delta & 0 & 0 & 0 & 0 & 0 & 0 & 0 & 0 & 0 & 0 \\
     -\frac{2\pi}{k}\delta & 0 & 0 & 0 & 0 & 0 & 0 & 0 & 0 & 0 & 0 & 0 \\
    0 & 0 & 0 & -\frac{2\pi}{3k} \delta'  & 0 & \frac{2\pi}{3k}\delta  & 0 & 0 & 0 & 0 & 0 & 0 \\
    0 & 0 &-\frac{2\pi}{3k}\delta'  & 0 & \frac{2\pi}{3k}\delta  & 0 & 0 & 0 & 0 & 0 & 0 & 0 \\
    0 & 0 & 0 &  -\frac{2\pi}{k}\delta & 0 & 0 & 0 & 0 & 0 & 0 & 0 & 0 \\
    0 & 0 &  -\frac{2\pi}{k}\delta & 0 & 0 & 0 & 0 & 0 & 0 & 0 & 0 & 0
  \end{pmatrix}.
\end{multline}
with
\begin{align*}
[C^{33}]^{-1}&=-\frac{4\pi^2}{k^2}[ (\cJ^1)' \delta +2 \cJ^1\delta'],\\
[C^{34}]^{-1}&=[C^{43}]^{-1}=\frac{2\pi}{3k}\delta'''-\frac{4\pi^2}{k^2}[ (\cP^1)' \delta +2 \cP^1\delta'],\\
[C^{36}]^{-1}&=-[C^{63}]^{-1}=[C^{45}]^{-1}=-[C^{54}]^{-1}=-\frac{2\pi}{3k}\delta''+\frac{4\pi^2}{k^2}\cP^1\delta.
\end{align*}
Furthermore, in order to compute the brackets of the densities with the constraints $\Phi$, the following relations are useful
\begin{align}
  \{\cM(\phi),\mathcal{P}^A(\phi')\}&=0,\\
  \{\cM(\phi),\mathcal{J}^A(\phi')\}&=-\cP^A(\phi)\partial_\phi\delta(\phi-\phi'),\\
  \{\cJ(\phi),\mathcal{P}^A(\phi')\}&=\cP^A(\phi)\partial_\phi\delta(\phi-\phi'),\\
  \{\cJ(\phi),\mathcal{J}^A(\phi')\}&=\cJ^A(\phi)\partial_\phi\delta(\phi-\phi'),\\
  \{\cW(\phi),\mathcal{P}^A(\phi')\}&=0,\\
  \{\cW(\phi),\mathcal{J}^A(\phi')\}&=-\frac{2\pi}{k}d^A_{\;\;BC}\cP^B(\phi)\cP^C(\phi)\partial_\phi\delta(\phi-\phi'),\\
  \{\cV(\phi),\mathcal{P}^A(\phi')\}&=\frac{2\pi}{k}d^A_{\;\;BC}\cP^B(\phi)\cP^C(\phi)\partial_\phi\delta(\phi-\phi'),\\
  \{\cV(\phi),\mathcal{J}^A(\phi')\}&=\frac{4\pi}{k}d^A_{\;\;BC}\cP^B(\phi)\cJ^C(\phi)\partial_\phi\delta(\phi-\phi').
\end{align}
Finally, $bms_3$ algebra \eqref{algD1} in terms of modes \eqref{modes} gives,
\begin{equation}\label{ss14}
\begin{split}
i\{P_n,P_m\}^{*}&=0,\\
i\{J_n,J_m\}^{*}&=(n-m)J_{n+m},\\
i\{J_n,P_m\}^{*}&=(n-m)P_{n+m}+kn^3\delta_{m+n},
\end{split}
\end{equation}
relations \eqref{algD2} become
\begin{equation}\label{ss16}
\begin{split}
i\{P_n,W_m\}^{*}&=0,\\
i\{J_n,W_m\}^{*}&=(2n-m)W_{n+m},\\
i\{P_n,V_m\}^{*}&=(2n-m)W_{n+m},\\
i\{J_n,V_m\}^{*}&=(2n-m)V_{n+m}.
\end{split}
\end{equation}
and \eqref{algD3} is expressed as
\begin{equation}\label{ss17}
\begin{split}
i\{W_n,W_m\}^{*}&=0,\\
i\{W_n,V_m\}^{*}&=\frac{1}{3}\left[\frac{8}{k}(n-m)\sum_{j=-\infty}^{\infty}P_jP_{n+m-j}+(n-m)(2n^2+2m^2-mn)P_{m+n}+kn^5\delta_{m+n}\right],\\
i\{V_n,V_m\}^{*}&=\frac{1}{3}\left[\frac{16}{k}(n-m)\sum_{j=-\infty}^{\infty}P_jJ_{n+m-j}+(n-m)(2n^2+2m^2-mn)J_{m+n}\right].
\end{split}
\end{equation}
Although algebra \eqref{ss14}, \eqref{ss16} and \eqref{ss17} is not, strictly speaking, a Lie algebra, it does fulfill Jacobi identity. 

\bibliography{/Users/hernangonzalez/Dropbox/boundaryW/BibHG}

\end{document}